\documentstyle[12pt,epsfig]{article}
\begin{document}

\begin{center}
{\Large \bf
ABOUT SOME RELATIVISTIC EFFECTS IN THE $A(\vec d,p)X$ REACTION
WITH THE PROTON EMISSION IN A FORWARD DIRECTION  }  \\
\vspace{9mm}

{\large
L.S.~Azhgirey$^1$,
N.P.~Yudin$^2$,
\vspace{9mm}

{\it $^1$ JINR, 141980 Dubna, Moscow Region, Russia } \\
{\it $^2$ Moscow State University, 119899 Moscow, Russia}  \\
}
\end{center}
\vspace{5mm}

\begin{abstract}
In a nonrelativistic quantum theory the longitudinal and
transversal components of the internal momentum in the deuteron
wave function (DWF) are bound together as they have to form the
well known superposition of the S- and D-waves. In a relativistic
case --- at least in light cone dynamics --- the dependence of the
DWF on the longitudinal and transversal momenta can by considerably
changed, because these components turn out in large part to be
bound in another way. Previously such a possibility was pointed
out by Blankenbecler et al.  Later Karmanov et al. have developed a
quantitative approach that can be used to describe this situation.
Here we present some results of calculations of the tensor
analyzing power of $(d,p)$ reaction at relativistic deuteron
momenta using Karmanov's DWF. The momentum behaviour of the tensor
analyzing power obtained is compared with that calculated with
standard DWFs.
\end{abstract}
\vskip 6mm

     In recent years attempts to gain some insight into a
manifestation of the structure of the relativistic deuteron
in nuclear interactions are concerned mainly with an
investigation of spin phenomena in reactions initiated by
relativistic polarized deuterons:
\begin{eqnarray}
  \vec{d} + p & \rightarrow & p + d,     \\
  \vec{d} + p & \rightarrow & p + p + n, \\
  \vec{d} + p(A) & \rightarrow & p + X,
\end{eqnarray}
where detected protons are emitted in forward directions
in the lab. syst. The simplest mechanism of these processes can be
represented schematically by the diagram shown in Fig. 1.
When protons are detected at non-zero angles, they may have
a rather high transversal component of the momentum.
At usually considered mechanisms these reactions are concerned
with the deuteron wave function.

     Already upon the first measurements of the tensor analyzing
power $T_{20}$ of reaction (3) at 9 GeV/$c$ with the emission of
protons at 0$^\circ$ [1-5] a significant discrepancy between the
values calculated in the relativistic impulse approximation and
experimental ones has come to light. A comparison of currently
available experimental data on $A_{yy}(0^\circ)$ (recall, that
at $0^\circ$ $A_{yy} = - T_{20}/\sqrt{2}$) with calculations in
the frame of the impulse approximation with the standard DWFs
\cite{paris,bonn} is made in fig. 2. More recently, on the
measurement of the tensor analyzing power $A_{yy}$ of reaction (3)
at 9 GeV/$c$ with the emission of protons with large transverse
momenta \cite{large_t}, this discrepancy has been compounded.
\vspace*{5mm}

{\unitlength=0.9mm
\thicklines
\large
\begin{picture}(150,66)
\put(20,51){\line(1,0){30}}
\put(20,49){\line(1,0){30}}
\put(37,53){\line(2,-1){6}}
\put(37,47){\line(2,1){6}}
\put(50,50){\line(3,1){30}}
\put(70,59.5){\line(3,-1){4}}
\put(70.2,54){\line(1,1){4}}
\put(50,50){\line(1,-1){30}}
\put(63.5,34.3){\line(2,-1){4}}
\put(65.8,38.3){\line(1,-3){2}}
\put(80,20.5){\line(6,1){30}}         % \line(1,0)
\put(80,19.5){\line(6,-1){30}}         % \line(1,0)
\put(97,25.8){\line(2,-1){4}}              % (2,-1)
\put(97,21.8){\line(2,1){4}}               % (2,1)
\put(97,18.2){\line(2,-1){4}}              % (2,-1)
\put(97,14.2){\line(2,1){4}}               % (2.1)
\put(50,10){\line(3,1){30}}
\put(70,19.5){\line(3,-1){4}}
\put(70.2,14){\line(1,1){4}}
\put(50,50){\circle*{4}}
\put(80,20){\circle*{4}}
\put(25,54){$d$}
\put(70,63){$p^\prime$}
\put(73,33){$ N \, (R)$}
\put(51,6){$p$}
\put(120,20){$ \Biggr\}$}
\put(130,27){$d^\prime$, \hspace*{11mm} {\rm or}}
\put(130,20){$p^\prime + n$, \hspace*{2mm} {\rm or}}
\put(130,13){$p^\prime + X$}
\end{picture}}
\vspace*{7mm}
\begin{quotation}
\parindent=0mm
Fig. 1. {\sf The simplest diagram to describe reactions initiated by
relativistic deuterons with proton emission in a forward direction.
Along with the one-nucleon exchange ($N$) there also the exchange of
baryon resonances ($R$) can take place \cite{azh2}. }
\end{quotation}

The discrepancy between the theory and experiment in principle can be 
overcome by the addition of a P-wave into the deuteron wave function 
(DWF). True enough, the P-wave is conditioned by the different
mechanisms in a number of approaches. In the paper \cite{kobushkin}
the data on $T_{20}$ at 0$^\circ$ have been successfully described
taking account of the P-wave arising because of the production of
six-quark configuration that gives odd parity resonances at their
fragmentation to the baryon channel. At the same time the
calculations on the basis of the Bethe-Salpeter equation give
the P-wave admixture that is inadequate to eliminate the
discrepancy with experiment \cite{kaptari}.
In relation to the reaction (1) it is notable that taking account
of an admixture of the odd parity baryon resonances to the DWF
allows the $T_{20}$ data at 0$^\circ$ to be better
described than within the usual one-nucleon exchange
approximation \cite{azh2}.

     In this report we would like to point out one more factor
related to the wave function of a fast moving deuteron. These
are the rotational properties of the relativistic DWF in the
light cone (LC) dynamics. \cite{lcd}. Let us consider shortly the
calculations in the frame of this approach.

     A general expression for the parameter $T_{20}$ has
the form
\begin{equation}
  T_{20} = \frac{\sum_{M,M^\prime}
           Sp\{\psi_M t_{20} \psi_{M^\prime}^{\dagger}\}}
           {(1/3) Sp\{\psi_M \psi_{M^\prime}^{ \dagger}\}},
\end{equation}
where the operator $t_{20}$ is defined by expression
\begin{equation}
  <m^\prime\ |\ t_{20}\ |\ m > =
  (-1)^{j-m^\prime} <1\ m\ 1\ -m^\prime\ |\ 2\ 0 >,
\end{equation}
$\psi_M$ is DWF with spin $J = 1$ and its projections on the
quantization axis $M = 0, \pm 1$, $\psi^\dagger$ is the Hermitian
conjugate function, and $<1\ m\ 1\ -m^\prime\ |\ 2\ 0 >$ is
a Clebsh-Gordan coefficient. In order to calculate $T_{20}$
we should know $\psi_M$ with different projections $M$.
However $\psi_M$ in LC dynamics has several unusual properties.
The first one consists in "strange" relations of $\psi_M$ with
different $M$. Let us suppose that we have $\psi_M$ with
$M = 0$. Then $\psi_M$ with $M = +1$ should be obtained from
$\psi_M$ with $M = 0$ by the action of the standard combination
of operators $J_x,\ J_y$:
\begin{equation}
\psi_{M=1} \sim (J_x + iJ_y) \psi_{M=0}.
\end{equation}
But in LC dynamics these operators of angular momentum are
dependent on the interaction, and to calculate the sum in the
expression for $T_{20}$ the Schr\"odinger equation for all
components of $\psi_M$ has to be solved independently.
Of course it is an objectionable problem. The dependence of
angular momentum on interaction involves difficulties in
forming states with definite angular momentum.

     The second complication centres around the dependence of
of the wave function on variables. Indeed, the wave function
of two particles is in general dependent on two variables,
\begin{equation}
\psi =  \psi(k_l, k_T),
\end{equation}
where $k_l$ and $k_T$ are longitudinal and transversal
components of momentum, respectively. In the nonrelativistic
physics $k_l$ and $k_T$ are rigidly bound together by the
necessity to form the usual superposition of $S$- and
$D$-waves in DWF, with the result that the nonrelativistic
DWF is dependent on the only variable $k$. In LC dynamics
this relation between $k_l$ and $k_T$ may be much weaker,
and in principle DWF can be dependent on two variables.
Attention to such a situation was first called by Schmidt
and Blankenbecler \cite{schmidt}.

     These difficulties of the LC-dynamics were to a great
extent overcame by Karmanov et al. \cite{karmanov,karm2}.
At first sight the approach of Karmanov complicates the
situation since there is introduced an additional dynamical
variable $\omega^\mu$ --- a four-vector of the normal to the
chosen light cone surface. But this new variable permits
the necessity dealing with operators of the angular momentum
dependent on the interaction to be excluded: it turns out
that the rotation of the light cone surface is equivalent to
the interaction dependent rotation of the system. Therefore
states with the definite angular momentum become formally
independent on the interaction.

     The dependence on the interaction does not completely
disappear, it arises in another context: there are two many
states with the same angular momentum and the role of the
interaction now is to choose the right combination among
these states. But this new problem is much more easier
than the problem of solving the Schr\"odinger equation.
The difficulties existing in Karmanov's approach were
overcame in \cite{karmanov,karm2}, where the parameters of the
relativistic DWF were given. This function was used in our
calculations of the tensor analysing power $A_{yy}$
of the deuteron breakup reaction (3).

     The DWF, now the function of nucleon spins, an internal
deuteron momentum, and an orientation of the quantization plane,
has the form:
\begin{equation}
  \Psi^M_{\sigma_2 \sigma_1} = w^\star_{\sigma_2}
  \psi^M(\vec{k}, \vec{n}) \sigma_y w_{\sigma_1},
\end{equation}
where
\begin{eqnarray}
  \vec{\psi}(\vec{k}, \vec{n}) = f_1\frac{1}{\sqrt{2}}\vec{\sigma} +
  f_2\frac{1}{2}[\frac{3}{\vec{k}^2}\vec{k}(\vec{k} \vec{n} - \vec{\sigma}] +
  f_3\frac{1}{2}[3\vec{n}(\vec{n} \vec{\sigma}) - \vec{\sigma}] +  \\
  f_4\frac{1}{2k}(3\vec{k}(\vec{n} \vec{\sigma}) +
     3\vec{n}(\vec{k} \vec{\sigma}) -
  2(\vec{k} \vec{n})\vec{\sigma}) +
  f_5\sqrt{\frac{3}{2}} \frac{i}{k} (\vec{k} \times \vec{n}) +
  f_6\frac{\sqrt{3}}{2k}((\vec{k} \times \vec{n}) \times \vec{\sigma}).
  \nonumber
\end{eqnarray}
Here $\vec{n}$ is the normal to the light front surface, $\vec{k}$ is
the momentum of nucleons in deuteron in their rest frame,
$\vec{\sigma}$ are the Pauli matrices, $w_{\sigma_1 (\sigma_2)}$
are the spin functions of nonrelativistic nucleons, and
$f_{1,...,6}$ are the invariant about rotations functions of
the kinematical variables, that define the deuteron state.

%\newpage
\begin{figure}[h]
\includegraphics[width=145mm]{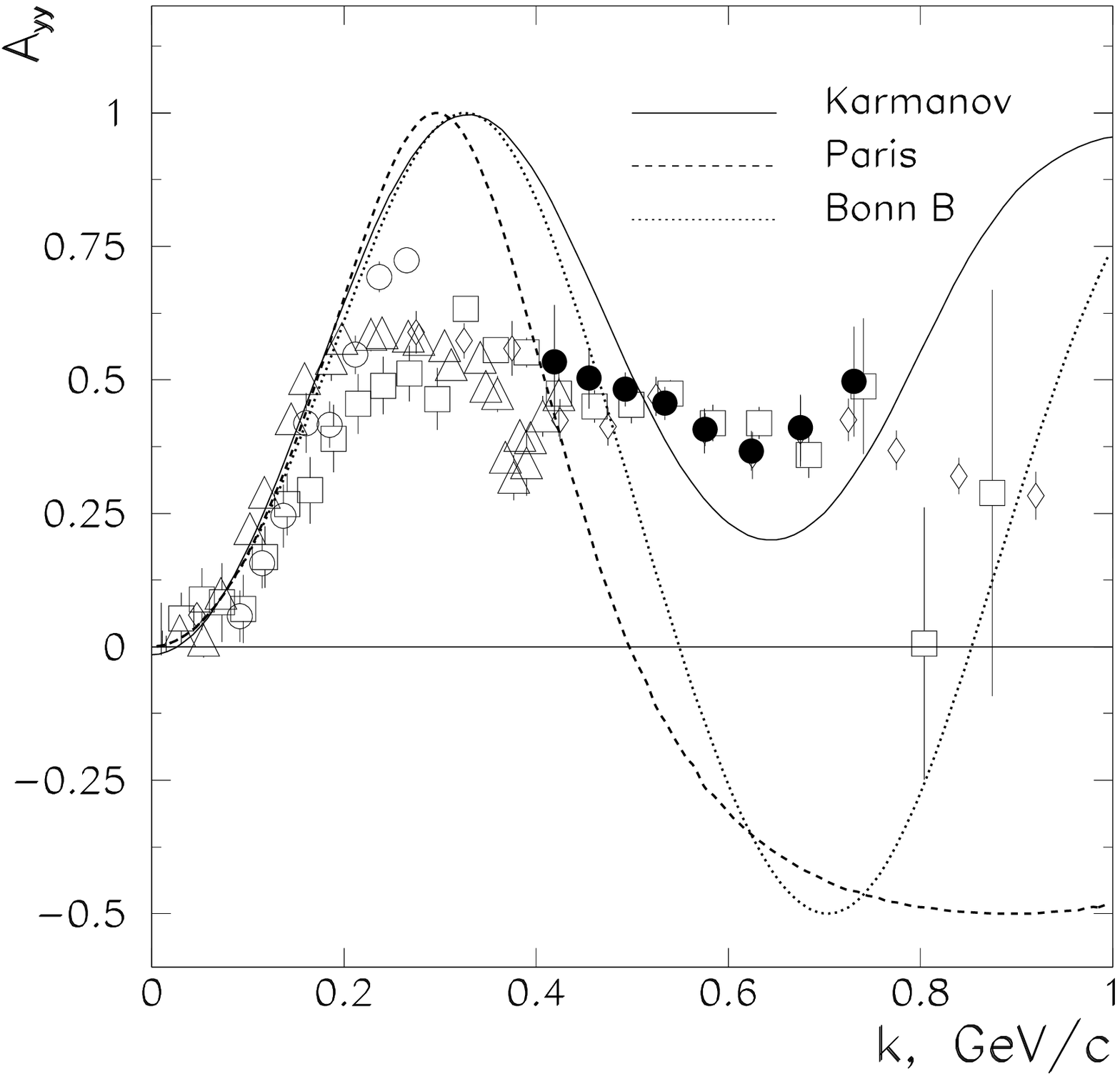}
\end{figure}
\begin{quotation}
\parindent=0pt
Fig. 2.
{\sl Tensor analyzing power $A_{yy}$ of the deuteron breakup with
the emission of protons at 0$^\circ$ vs internal momentum $k$.
Experimental data: triangles --- \cite{perdrisat,punjabi},
squares --- \cite{ableev}, diamonds --- \cite{aono}, empty
circles --- \cite{azhgirey} ($H$ target), full circles ---
\cite{azhgirey} ($C$ target). Calculations are made in relativistic
impulse approximation with Paris \cite{paris} (broken curve),
Bonn \cite{bonn} (dotted curve) and Karmanov's \cite{karmanov} DWFs. }
\end{quotation}

     Previously we described $dp \rightarrow pX$ and
$dp \rightarrow pd$ reactions \cite{azh2} in the
coordinate system where the $z$ axis was aligned with the
beam direction. For the wave function (8), too, let us
assume that $\vec{n}$ is directed along $z$ axis, i.e.
$\vec{n} = (0, 0, 1)$.

     The expression (4) may be written in the form:
\begin{equation}
  T_{20} = \sqrt{\frac{3}{2}}
           \frac{[-3Sp(\psi_z \psi_z^\dagger) +
           Sp(\vec{\psi} \vec{\psi}^\dagger)]}
           {Sp(\vec{\psi} \vec{\psi}^\dagger)},
\end{equation}
where we went from spherical notations to the vector ones.
For the traces we have:
\begin{eqnarray}
  Sp(\vec{\psi} \vec{\psi}^\dagger) & = & 3[f_1^2 + f_2^2 +
  (3z^2 - 1)f_2f_3 + f_3^2 + 4zf_4(f_2 + f_3) +         \nonumber \\
  & &  f_4^2(z^2 + 3) + (1 - z^2)f_5^2 + (1 - z^2)f_6^2],  \\
  Sp(\psi_z \psi_z^\dagger) & = & f_1^2 + \sqrt{2}[3(kz)^2 - 1]f_1f_2 +
  2\sqrt{2}f_1f_3 + 4\sqrt{2}kzf_1f_4 +       \nonumber  \\
  & &  \frac{1}{2}[3(kz)^2 + 1]f_2^2 + 2[3(kz)^2 - 1]f_2f_3 +
  kz[3(kz)^2 + 5]f_2f_4 +               \nonumber  \\
  & &  2f_3^2 + 8kzf_3f_4 + 3sqrt{3}kz[1 - (kz)^2]f_2f_6 +
  3sqrt{3}[1 - (kz)^2]f_4f_6 +       \nonumber  \\
  & &  \{[\frac{9}{2}[(kz)^2 + 1] - (kz)^2\}f_4^2 +
  \frac{3}{2}[1 - (kz)^2]f_6^2,        \nonumber
\end{eqnarray}
where $z = cos \theta$.

     The results of calculations of $A_{yy}(0^\circ)$ with
the wave function (9) are shown in fig. 2 by the solid curve.
It is seen that as opposed to the calculations with the standard
nonrelativistic DWFs \cite{paris,bonn} the solid curve does not
cross the horizontal axes and is in reasonable good agreement
with experimental data in the region of $k$ from 0.4 to 0.8 GeV/$c$.
This result in our opinion is due to the fact that Karmanov's
model establishes a new link between $k_l$ and $k_T$ that is
different from those of the S- and D-wave superposition in
nonrelativistic DWF. Similar effect was discussed in \cite{karm2}
on the example of Wick-Cutcosky model, where it was shown in
the clear form that a $S$-wave two-particle system becomes
dependent on the angle in the LC dynamics. In other words it
is a manifestation of the intimate connection between the
internal motion and the motion of the system as a whole.

     This work was supported in part by the Russian Foundation for
Fundamental Research (grant No. 01-02-17299).


\begin{thebibliography}{99}
  \bibitem{perdrisat}
C.F.Perdrisat {\it et al.}, Phys.Rev.Lett. {\bf 59}, 2840 (1987).
  \bibitem{punjabi}
V.Punjabi {\it et al.}, Phys.Rev. C{\bf 39}, 608 (1989).
  \bibitem{ableev}
V.G.Ableev {\it et al.}, JINR Rapid Comm. No.4[43]-90, 5 (1990).
  \bibitem{aono}
T.Aono {\it et al.}, Phys.Rev.Lett. {\bf 74}, 4997 (1995).
  \bibitem{azhgirey}
L.S.~Azhgirey {\it et al.}, Phys.Lett. B{\bf 387}, 37 (1996).
  \bibitem{paris}
M.Lacombe {\it et al.}, Phys.Rev. C{\bf 21}, 861 (1980).
  \bibitem{bonn}
R.Machleidt {\it et al.}, Phys.Rep. {\bf 149}, 1 (1987).
  \bibitem{large_t}
S.V.Afanasiev {\it et al.}, Phys.Lett. B{\bf 434}, 21 (1998).
  \bibitem{kobushkin}
A.P.Kobushkin, Phys.Lett. B{\bf 421}, 53 (1998).
  \bibitem{kaptari}
L.Kaptari {\it et al.}, Phys.Lett. B{\bf 351}, 400 (1995).
  \bibitem{azh2}
L.S.Azhgirey and N.P.Yudin, Yad.Fiz. {\bf 63}, 2280 (2000)
(Phys.Atom.Nucl. {\bf 63}, 2184 (2000)).
  \bibitem{lcd}
P.A.M.~Dirac, Rev.Mod.Phys. {\bf 21}, 392 (1949);  \\
S.Weinberg, Phys.Rev. {\bf 150}, 1313 (1966);     \\
L.L.Frankfurt and M.I.Strikman, Phys.Rep. {\bf 76}, 215 (1981).
  \bibitem{schmidt}
I.A.Schmidt and R.Blankenbecler, Phys.Rev. D{\bf 15}, 3321 (1977); \\
Ch.-Y.Wong and R.Blankenbecler, Phys.Rev. C{\bf 22}, 2431 (1980); \\
M.Chemtob {\it et al.}, Nucl.Phys. A{\bf 314}, 387 (1979).
  \bibitem{karmanov}
V.A.Karmanov and A.V.Smirnov, Nucl. Phys. A {\bf 575}, 520 (1994); \\
J.Carbonell {\it et al.}, Phys.Rep. {\bf 300}, 215 (1998).
  \bibitem{karm2}
J.Carbonell and V.A.Karmanov, Nucl. Phys. A {\bf 581}, 625 (1994).

\end{thebibliography}
\end{document}